\documentclass[letter]{aa}
\usepackage{txfonts}
\usepackage{graphicx}
\usepackage{rotating}
\usepackage{times}
\usepackage{amsmath}
\usepackage{multirow}

\usepackage{natbib,twoopt}
\usepackage[breaklinks=true]{hyperref} 
\bibpunct{(}{)}{;}{a}{}{,}             
\makeatletter
  \newcommandtwoopt{\citeads}[3][][]{\href{http://adsabs.harvard.edu/abs/#3}%
    {\def\hyper@linkstart##1##2{}%
     \let\hyper@linkend\@empty\citealp[#1][#2]{#3}}}
  \newcommandtwoopt{\citepads}[3][][]{\href{http://adsabs.harvard.edu/abs/#3}%
    {\def\hyper@linkstart##1##2{}%
     \let\hyper@linkend\@empty\citep[#1][#2]{#3}}}
  \newcommandtwoopt{\citetads}[3][][]{\href{http://adsabs.harvard.edu/abs/#3}%
    {\def\hyper@linkstart##1##2{}%
     \let\hyper@linkend\@empty\citet[#1][#2]{#3}}}
  \newcommandtwoopt{\citeyearads}[3][][]%
    {\href{http://adsabs.harvard.edu/abs/#3}
    {\def\hyper@linkstart##1##2{}%
     \let\hyper@linkend\@empty\citeyear[#1][#2]{#3}}}
\makeatother

\newcommand{\Lsun} {L$_\odot$}

\newcommand{\Rout} {R$_{\rm out}$}

\newcommand{\micron} {\,$\mu$m}
\newcommand{\simless}{\mathbin{\lower 3pt\hbox
      {$\rlap{\raise 5pt\hbox{$\char'074$}}\mathchar"7218$}}} 
\newcommand{\simgreat}{\mathbin{\lower 3pt\hbox
     {$\rlap{\raise 5pt\hbox{$\char'076$}}\mathchar"7218$}}} 
\newcommand{\Lprime}{$L^\prime$}

\begin{document}
\title{GG~Tau: the fifth element\thanks{Based on ESO programs 090.C-0339(A,F)}} 
\author{
 E.~Di Folco \inst{1,2},
 A.~Dutrey \inst{1,2},
J.-B.~Le Bouquin \inst{3},
S.~Lacour\inst{4},
J.-P.~Berger\inst{5},
R.~K\"ohler \inst{6},
S.~Guilloteau \inst{1,2},
V.~Pi\'etu\inst{7},
J.~Bary\inst{8}, 
T.~Beck\inst{9},
H.~Beust\inst{3},
E.~Pantin\inst{10}
 }
\institute{
Univ. Bordeaux, Laboratoire d'Astrophysique de Bordeaux, UMR 5804, F-33270, Floirac, France
\and{}
CNRS, LAB, UMR 5804, F-33270 Floirac, France\\
  \email{emmanuel.difolco@obs.u-bordeaux1.fr}
 \and{}
UJF-Grenoble 1/CNRS-INSU, Institut de Plan\'etologie et d'Astrophysique de Grenoble UMR 5274, F-38041, Grenoble, France
\and{}
LESIA, CNRS/UMR-8109, Observatoire de Paris, UPMC, Universit\'e Paris Diderot, 5 place J. Janssen, F-92195, Meudon, France
\and{}
European Southern Observatory, D-85748, Garching by M\"unchen, Germany
\and{}
Max-Planck-Institut f\"ur Astronomie, K\"onigstuhl 17, D-69117 Heidelberg, Germany
\and{}
IRAM, 300 rue de la piscine, F-38406 Saint-Martin d'H\`eres, France
\and{}
Department of Physics and Astronomy, Colgate University, 13 Oak Drive, Hamilton, NY 13346, USA
\and{}
Space Telescope Science Institute, 3700 San Martin Dr. Baltimore, MD 21218, USA; tbeck@stsci.edu, lubow@stsci.edu
\and{}
Laboratoire AIM, CEA/DSM - CNRS - Universit\'e Paris Diderot, IRFU/SAP, F-91191, Gif-sur-Yvette, France,
}


\date{Received 19 February 2014, Accepted 10 March 2014} %
\authorrunning{Di~Folco et al.} %
\titlerunning{GG Tau: the fifth element}

\abstract
{We aim at unveiling the observational imprint of physical mechanisms 
that govern planetary formation in young, multiple systems. 
In particular, we investigate the impact of tidal truncation on the inner circumstellar disks. 
We observed the emblematic system GG~Tau at high-angular resolution: 
a hierarchical quadruple system composed of low-mass T~Tauri binary stars 
surrounded by a well-studied, massive circumbinary disk in Keplerian rotation. 
We used the near-IR 4-telescope combiner PIONIER on the VLTI and 
sparse-aperture-masking techniques on VLT/NaCo to probe this proto-planetary system at sub-au scales. 
We report the discovery of a significant closure-phase signal in $H$ and $K_s$ bands 
that can be reproduced with an additional low-mass companion orbiting GG~Tau~Ab, 
at a (projected) separation $\rho = 31.7\pm 0.2$\,mas (4.4\,au) and $PA = 219.6 \pm 0.3^\circ$. 
This finding offers a simple explanation for several key questions in this system, 
including the missing-stellar-mass problem and the asymmetry of continuum emission 
from the inner dust disks observed at millimeter wavelengths. Composed of now five co-eval stars with 
$0.02 \le M_{\star} \le 0.7 $\,M$_{\odot}$, the quintuple system GG~Tau has become an ideal test case 
to constrain stellar evolution models at young ages (few $10^6$\,yr).
}
\keywords{Stars: binaries: close - Planetary systems: proto-planetary disks  - Techniques: high angular resolution, interferometry}

\maketitle

\section{Introduction}
Planet formation is a common process that can occur in different environments.
While the first decade of planet searches has preferentially focused on single, solar-like host stars,
it has been more recently shown that a large proportion of extrasolar giant planets are born
in binary systems \citep{udry07}.
The recent diskoveries of transiting circumbinary planets
in close binary systems \citep[Kepler 16, 34,35, ][]{doyle11, welsh12},
as well as the planet candidate directly imaged around the
young low-mass binary 2MASS J0103 \citep{delorme13} have proven that planets
can also appear in a circumbinary disk, despite the strong dynamical mechanisms
that shape the disk and can rapidly clear out its inner region.
Stars in young binary systems are expected to be surrounded by
two inner disks, located inside the Roche lobes and an outer circumbinary ring or disk
outside the outer Lindblad resonances \citep[e.g., ][]{artymowicz94}.
Persistent signs of accretion in binary systems, as well as direct imaging
of residual gas in the inner region, demonstrate that gas and dust can flow
from the outer reservoir through this gravitationally unstable zone
to nurture inner circumstellar disks
(where planet formation may also occur), which otherwise would not survive.
Understanding how the inner disks are replenished is also important in the general context
of planetary system formation, since binary stars provide a scaled-up version
of a proto-planet environment in a circumstellar (CS) disk.
Finally, mutiple systems can provide essential clues in testing stellar evolution
models, as they provide a set of co-eval stars with different masses at a common
distance \citep[e.g,][]{white99}

In the past two decades, the young hierarchical quadruple 
system GG~Tau, composed of two low-mass binary systems, 
has been subject of many detailed studies. 
With its relatively massive (0.15\,M$_{\odot}$) and bright outer ring,
\object{GG~Tau~A} is one of the best known nearby (140\,pc) T~Tauri binaries,
with a 0.26\arcsec\ separation ($36$\,au on the sky plane).
The circumbinary disk ($R_{\rm in} =180$\,au) has been observed in thermal
dust emission \citep{dutrey94, guilloteau99,pietu11} and in scattered light
\citep{roddier96,silber00}, and is in Keplerian rotation \citep{dutrey94}.
The scattered-light images proved that the gravitationally unstable zone is not empty of dust.
Indirect evidence for gas flow from the ring towards the inner system(s) has been found
from $^{12}$CO J=2-1 gas image \citep{guilloteau01} and from near-IR H$_2$ transitions \citep{beck12}.
The warm H$_2$ gas may be heated by shocks, as material from the circumbinary ring is accreted
onto material close to the stars. The existence of inner CS disks
is independently attested by mm excess emission on GG~Tau~Aa \citep{pietu11},
strong H$_\alpha$ accretion signature separately detected around Aa and Ab,
[OI] line detection around Ab \citep{white99,hartigan03},
and 10$\mu$m silicate feature from hot grains in both Aa and Ab environments \citep{skemer11}.
We have recently undertaken a very high-spatial resolution
observing program of GG~Tau~A, from UV to mm wavelengths. 
In this letter, we report near-IR VLT interferometric observations of the inner region
of the GG~Tau~A system, where we detect a new component and
direct evidence for resolved circumstellar dust emission.
\begin{figure*}
\centering
\includegraphics[width=5.5cm,angle=0]{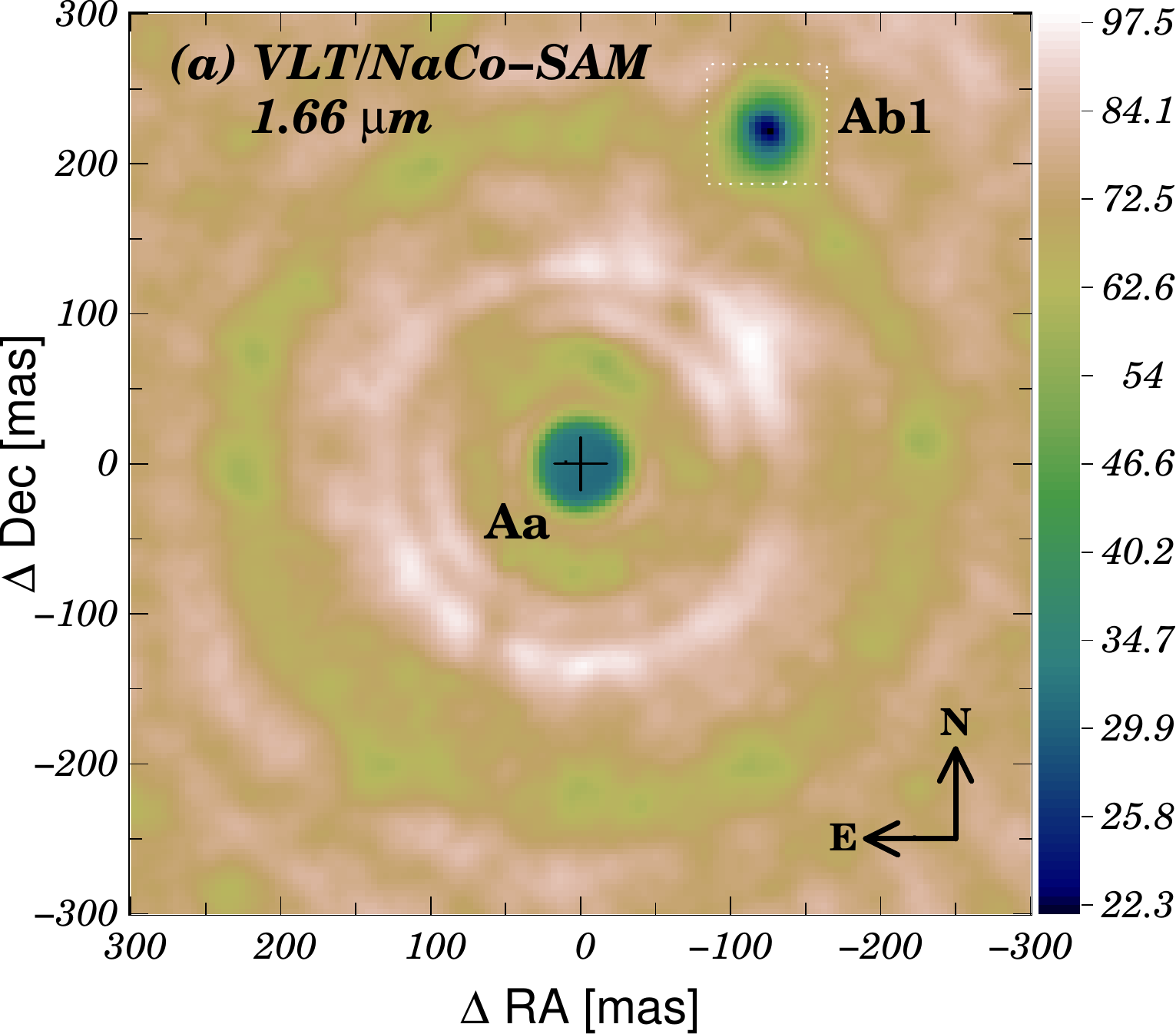}
\includegraphics[width=5.5cm,angle=0]{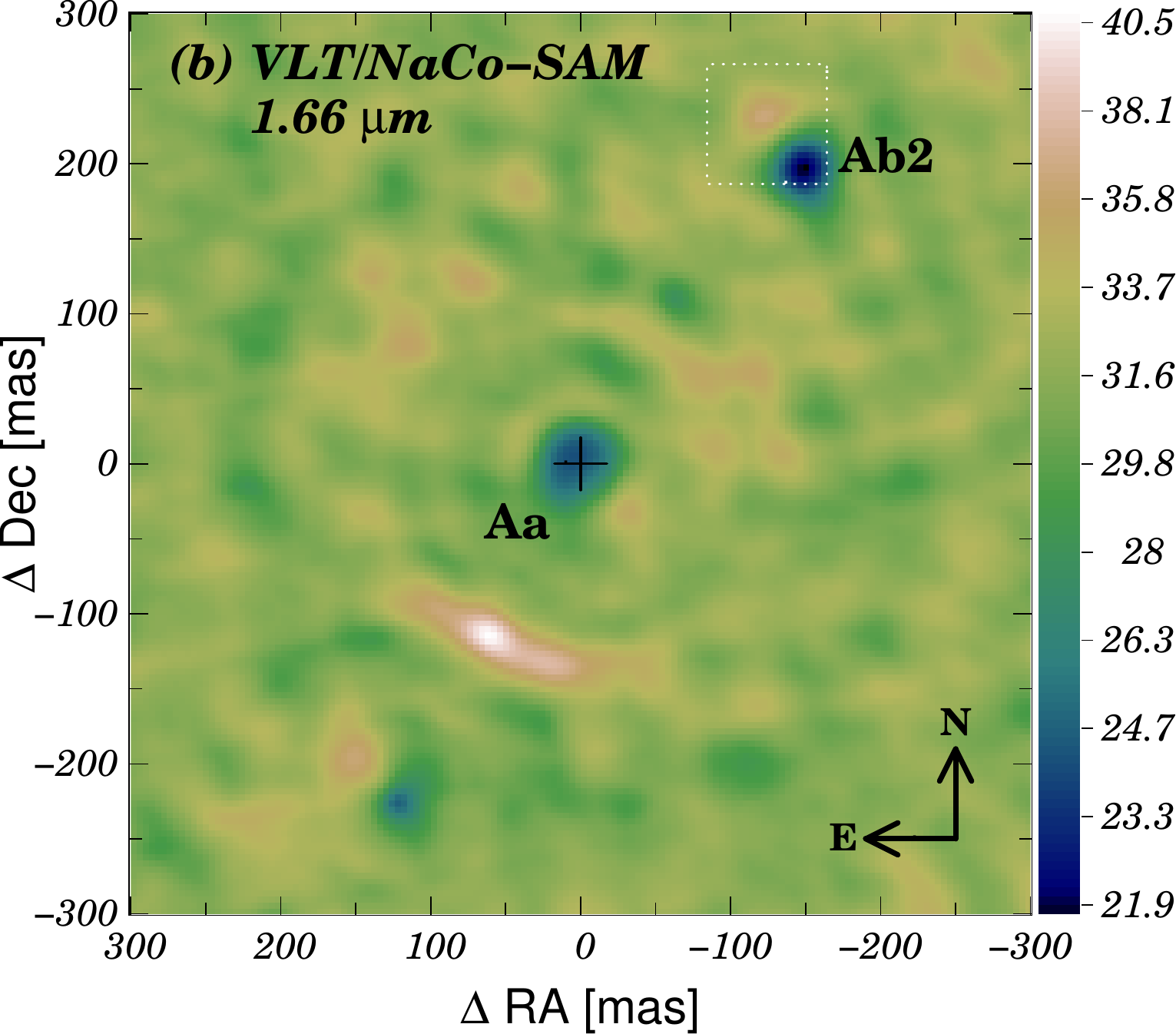}
\includegraphics[width=5.35cm,angle=0]{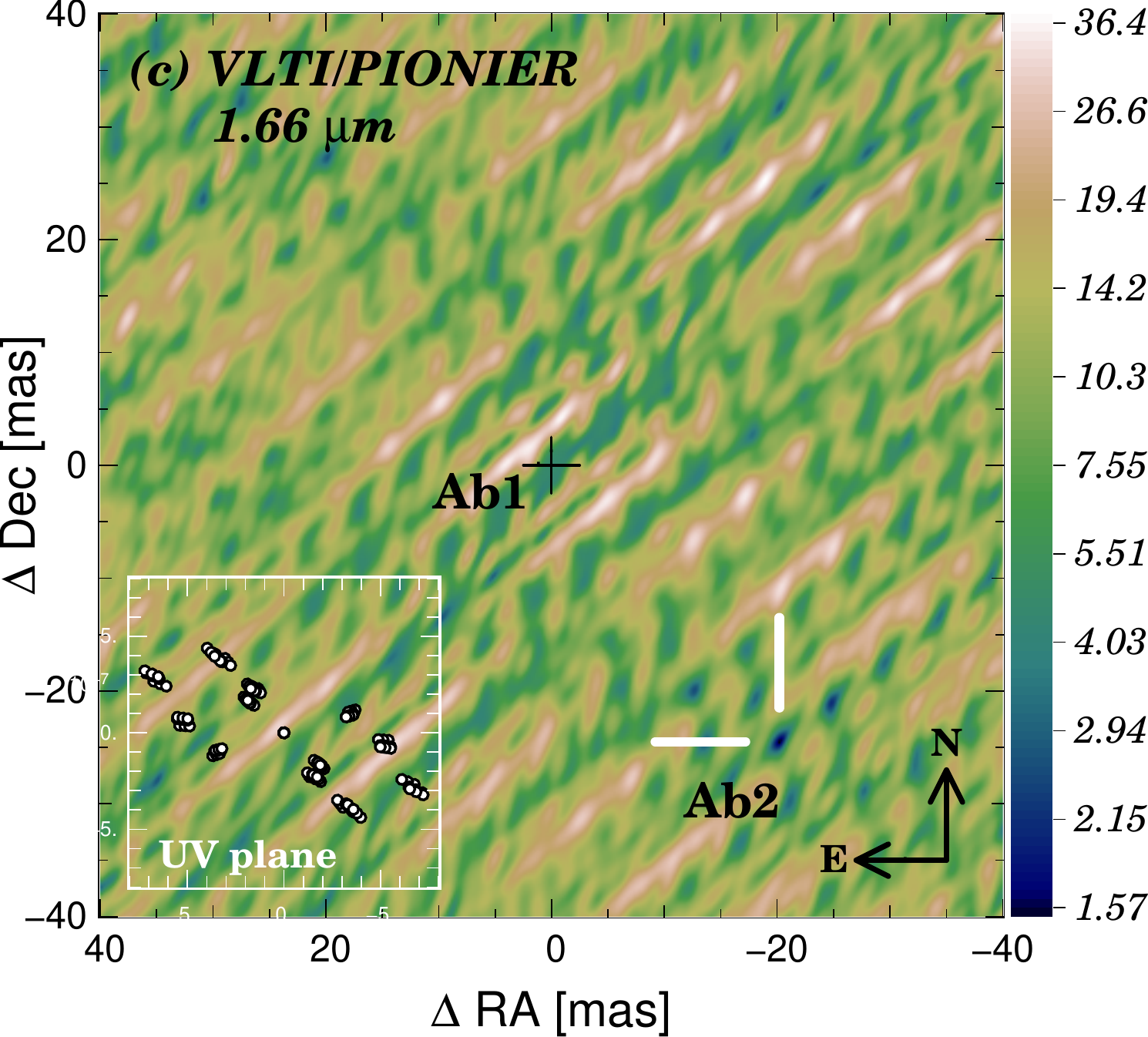}
\caption{Chi2 maps for: NACO-SAM $H$-band closure-phase data (6+7 Dec. 2012):
location of GG~Tau Ab1 (a) and Ab2 (b) around the primary star GG~Tau Aa, 
and for PIONIER (c) $H$-band closure-phase data for Ab2 location around Ab1, with VLTI $(u,v)$ sampling in the inset.
  }
  \label{fig_chi2maps}
\end{figure*}

\section{Observations and data analysis}
We observed the GG~Tau A system on 2012 October 30 with the VLTI,
using the four 8\,m Unit Telescopes (UTs) on baseline lengths between 32 and 140\,m,
and the PIONIER instrument \citep{lebouquin11} operating in the $H$ band 
($1.5-1.80$\micron, three spectral channels). 
Thanks to the combination of the MACAO adaptive optics system and the tiny interferometric
field of view (FOV) of VLTI /UTs ($FWHM\sim 41$\,mas in $H$ band),
we were able to separately point at the 0.26\arcsec\ binary GG~Tau~Aa (M0V) and Ab (M2V) with PIONIER. 
Seeing conditions were stable, with optical seeing values 
$\theta_{0}\sim0.6-0.9\arcsec$, and a coherence time in the range $\tau_{0}\sim3-5$\,ms. 
We performed interleaved observations of two calibrator sources of known diameter 
(HD28462, K1V, $\Theta_{\rm UD}^{H} = 0.169 \pm 0.012$\,mas, and 
HD285720, K4V, $\Theta_{\rm UD}^{H} = 0.148 \pm 0.011$\,mas). 
Four independent series of calibrated observations were acquired, 
with interferometric fringes simultaneously recorded on six  baselines, 
and reduced with the {\it pndrs} package \citep{lebouquin11}.

We also observed GG~Tau A on 2012 December 6 and 7
with the diffraction-limited imager VLT/NAOS-CONICA \citep{lenzen03}. 
We performed sparse-aperture-masking (SAM) observations on NaCo with 
a 7-hole mask \citep{tuthill00}. 
The mask at the pupil-plane blocks most of the light from the centered target
and resamples the primary mirror into a set of smaller subapertures that form a sparse
interferometric array with 21 baselines. 
The 0.26\arcsec\ close binary Aa--Ab is resolved in the NaCo field of view, 
but not by the individual subpupils. 
Measurements of closure-phases ($CP$) allow one to detect high-contrast companions 
and proved to be more efficient than classical AO-imaging 
within $\sim20-300$\,mas at $H$ and $K_s$-bands \citep{lacour11}.
We used a 27\,mas/pixel plate scale and high-cadence frame acquisition mode,
the science target itself being used for the IR wavefront sensing.
On Dec. 6, data were recorded with the $H$ and $K_s$ bands ($2.0-2.1$\micron) filters, 
while $H$ and \Lprime\ filters were used on Dec. 7. 
Atmospheric conditions were stable ($\theta_{0}\sim0.5-0.8\arcsec$, $\tau_{0}\sim3-6$\,ms).
The same two calibrator sources as for PIONIER were selected,
with a fast-switch pointing sequence. 
The observations were reduced using the Paris SAMP pipeline as described in \citet{lacour11}.

\section{Results}
\begin{figure}
\centering
\includegraphics[width=4.5cm,angle=0]{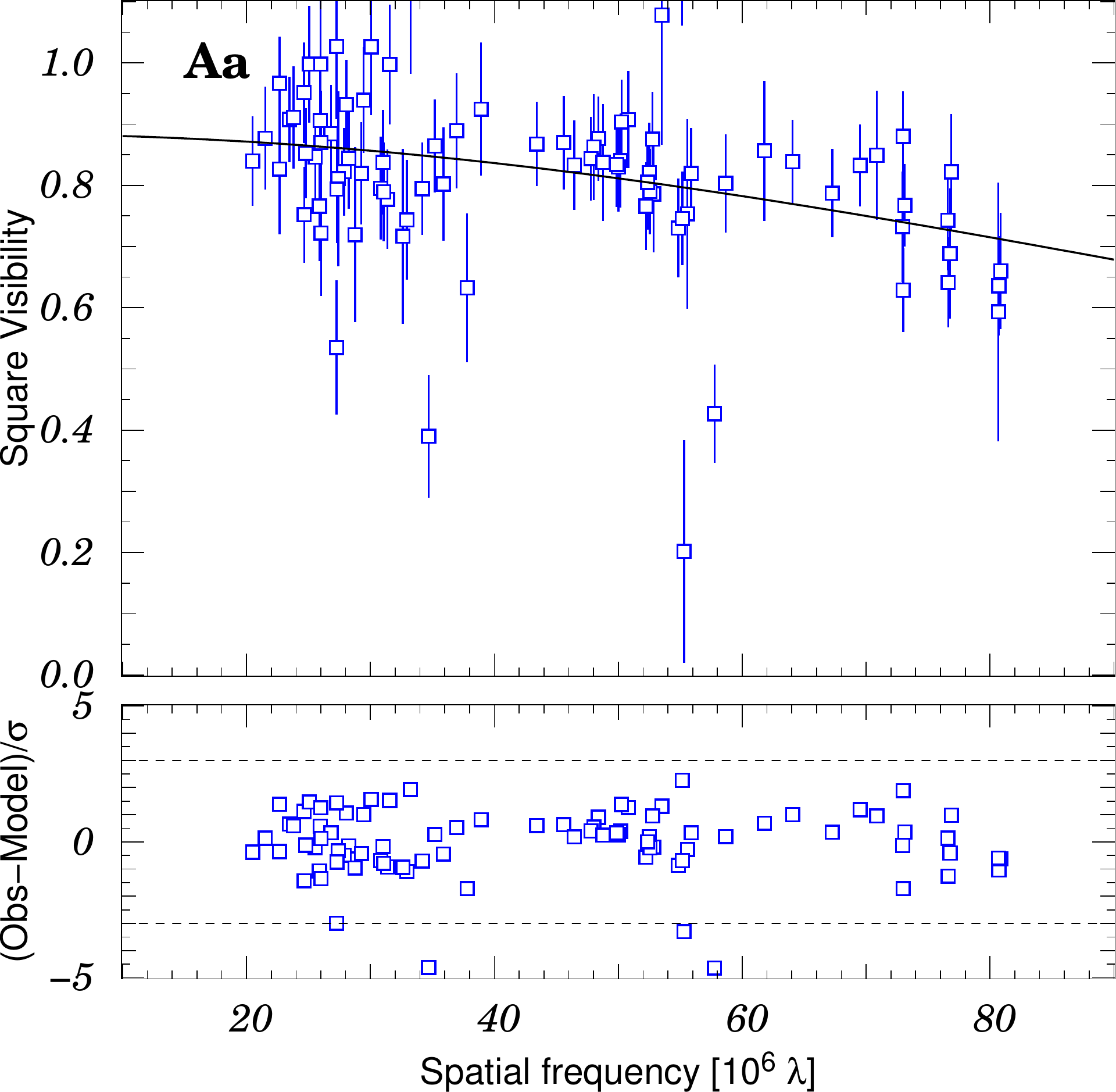}
\includegraphics[width=4.35cm,angle=0]{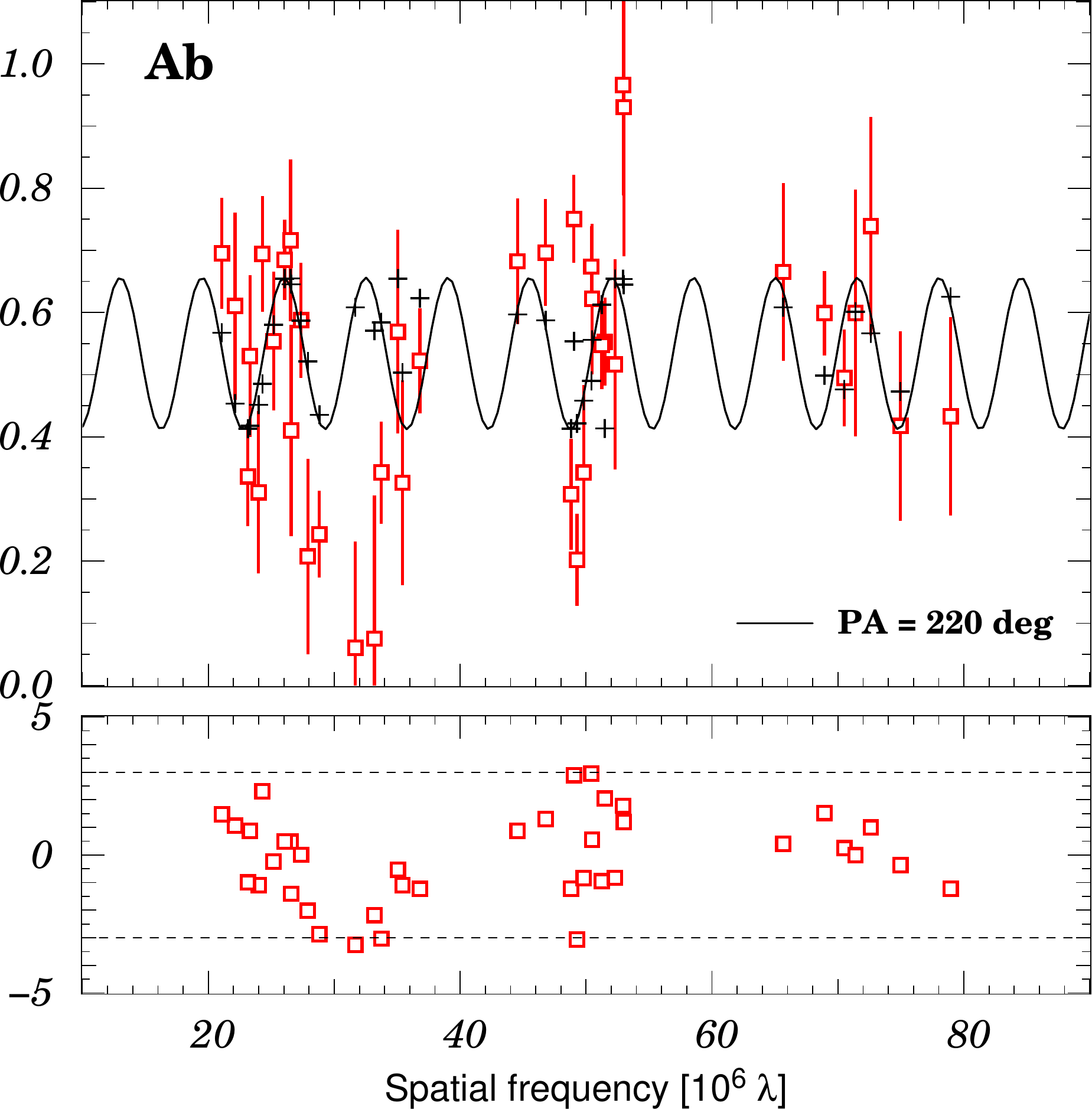}\\
\includegraphics[width=4.5cm,angle=0]{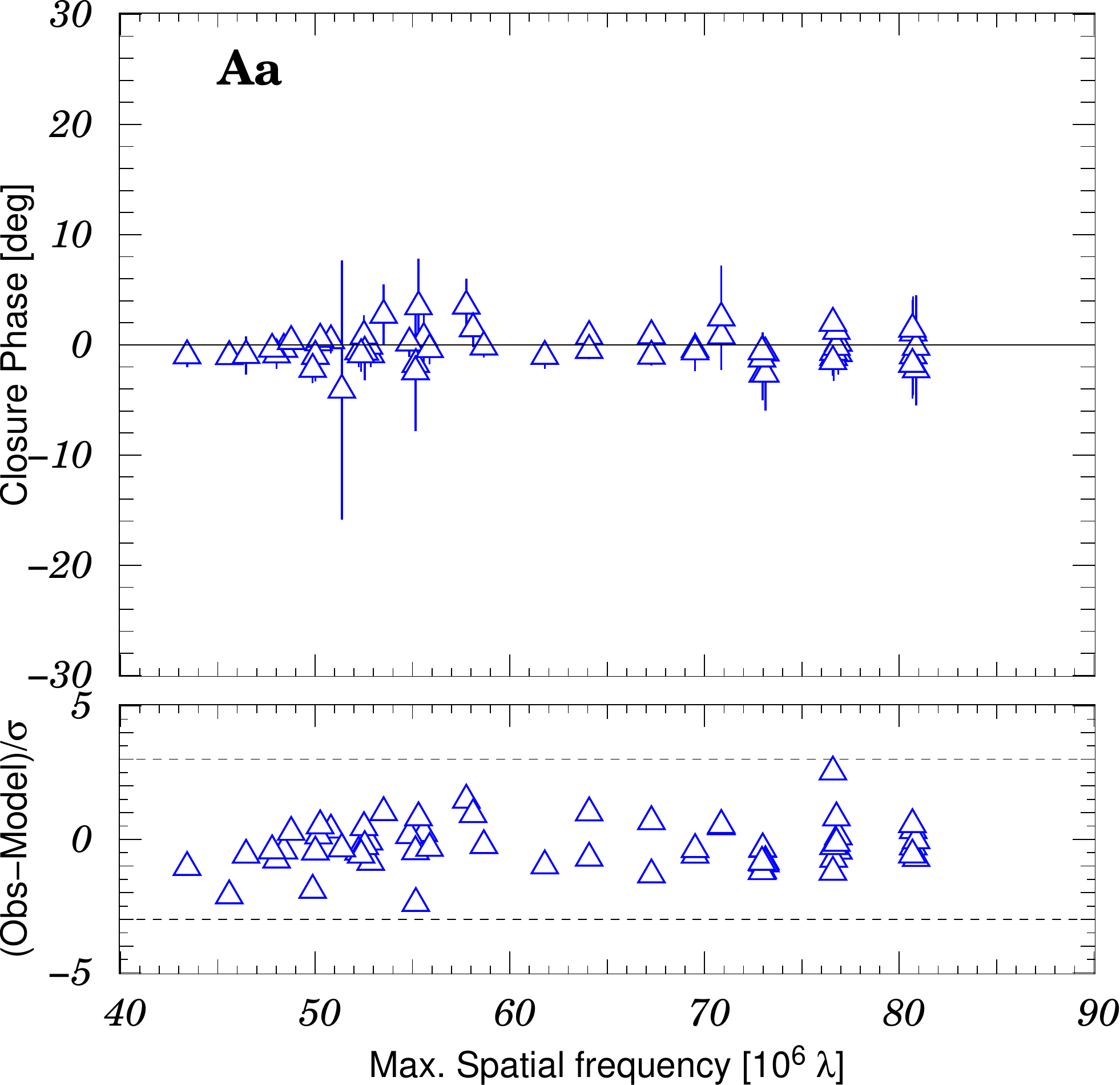}
\includegraphics[width=4.4cm,angle=0]{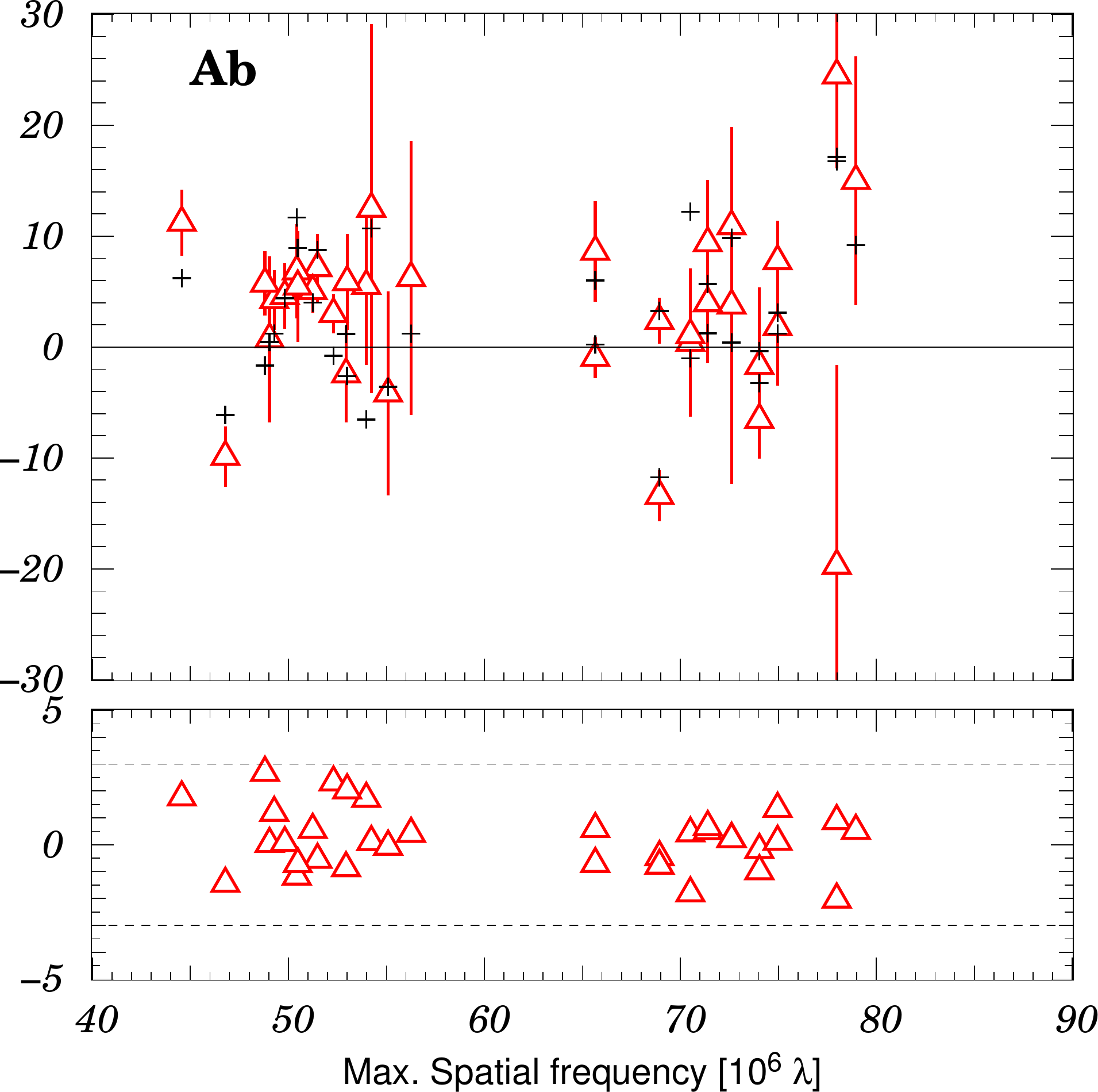}
\caption{VLTI/PIONIER square visibilities (squares, top panels) and closure phases (triangles, bottom panels) separately measured for GG~Tau Aa (left, blue) and Ab1+2(right). Solid lines or cross markers represent best-fit models: unresolved photosphere, halo and ring for Aa (model-2 in Table~\ref{table_interferometry}), and 2 unresolved photospheres and a halo for Ab (model-4). 
Below each plot we display the fit residuals.
  }
  \label{fig_v2+t3}
\end{figure}

The PIONIER measurement for GG~Tau~Aa is consistent with 
a marginally resolved, symmetric emission, as attested by the fringe visibility ($V^2$) 
and $CP$ functions displayed in Fig.\ref{fig_v2+t3}. 
Around the dimmer GG~Tau~Ab, we report the detection of a $CP$ signal as large as 30\,deg,
which reveals an asymmetric brightness distribution. 
We attribute this feature to a third companion (Ab2) in the main binary system GG~Tau~A. 
Square visibilities also show oscillation-like structures, 
but we chose to rely on the $CP$ values to derive the binary characteristics, 
because they are less affected by atmospheric phase fluctuations and known telescope vibrations. 
We used the LITPRO software \footnote{www.jmmc/litpro} \citep{tallon08}
to extract the Ab1--Ab2 binary flux ratio and astrometric position by independently 
fitting the PIONIER and NaCo-SAM data (Table~\ref{table_astrometry}). 
A simple model with two unresolved stars yields a good fit to the $CP$ values  
($\chi^2_{\rm red}$ =1.5 vs 7.1 for a single star), with a raw flux ratio of $8.1 \pm 2.1$ ($H$-band). 
We derive a separation of $31.7\pm 0.2$\,mas and a 
position angle PA$=219.6 \pm 0.3\deg$ (Fig.\ref{fig_chi2maps}). 
\footnote{Because of the limited $(u,v)$ coverage with VLTI, 
a second position for Ab2 is possible ($\rho = 32.5 \pm 0.2$\,mas, $PA=230.0 \pm 0.4\deg$), 
although less likely ($\chi_r^2 = 2.0$).}
Because the IR photo-center of Ab1--Ab2 is closer to Ab1 
and their separation represents about 30\,\% of the instrumental FOV (i.e., an 8\,m telescope Airy disk), 
we applied a correction for the FOV attenuation. 
The correction factor (based on a Gaussian-profile attenuation of $FWHM=41$\,mas, centered on the $K_s$-band centroid) 
amounts to $f_{\rm corr} = 0.53$. 
(The uncertainty on $f_{\rm corr}$ depends on the AO and tip-tilt stabilizer performance; 
we evaluate $\delta f_{\rm corr} \sim 0.2$ for a residual uncertainty 4\,mas rms on the position of the Ab1--Ab2 centroid).
The attenuation-corrected PIONIER flux ratio for a two-point source-model 
becomes $(F_{\rm Ab1}/F_{\rm Ab2})_{\rm corr} =4.3 \pm1.1 \pm 1.6$ (the second term is due to $\delta f_{\rm corr}$ only),
 that is $\Delta H \sim1-2$\,mag.

The two components of GG~TAu~A were simultaneously observed with NaCo. 
The $CP$ values are dominated by the Aa--Ab binary, and we first fit its position: 
the 256\,mas separation is consistent with long-term astrometric studies \citep{koehler11},
 and the flux ratios agree with literature data. 
We then added a third component Ab2 by fitting its location relative to Aa and its flux ratio. 
The Ab1--Ab2 close binary is well constrained by the NaCo data, and the three-component model always yields 
a better fit. The best-fit positions are consistent with each other at $H$ and $K_s$ bands, 
as well as with the VLTI estimate (although with larger uncertainties, see Table~\ref{table_astrometry}). 
We adopted an averaged separation of $31.6 \pm 3.3$\,mas and PA$=222.6\pm6.0\deg$ with NaCo.
In the less angularly resolved \Lprime\ band, we fixed the binary separation to its mean value and only 
fit the flux ratio ($2.6 \pm 0.4$). 
The mask subaperture sizes correspond to an equivalent pupil diameter of 1.2\,m, 
hence a 350\,mas point spread function (PSF) in $H$ band. 
To properly account for the resulting flux attenuation, we oversampled the Fourier plane by a factor of 3. 
The derived NaCo-SAM flux ratios are then $1.6\pm0.4$ and $2.4 \pm 0.3$ at $H$ and $K_s$ bands, respectively.
These values were used to estimate the spectral type of Ab2. 
Surprisingly, the inferred Ab1/Ab2 $H$-band flux ratio appears to be lower than our PIONIER estimate ($4.3 \pm 2.3$). 
This discrepancy is discussed in Appendix~\ref{Sec:innerdisk}. 

\section{Discussion}
\subsection{Nature of the detected component}\label{Sec:nature}
A non zero closure-phase indicates a non axisymmetric brightness distribution,
which may either be interpreted as the presence of a stellar companion,
or as a brightness asymmetry in the disk emission. Similar detections with SAM techniques
have recently been reported for a few other young sources with known protoplanetary disks 
\citep[e.g., T~Cha, ][]{huelamo11}. For these transitional disks,
it has been demonstrated that starlight scattered off the inner edge of the outer disk
(typically at $R\sim10$\,au) can equally well reproduce the detected closure-phase
as a (sub)stellar companion \citep{olofsson13,cieza13}.
However, the $CP$ amplitudes in these systems are much smaller than for GG~Tau~Ab 
($\lesssim 1\deg$, implying much higher IR contrasts $\Delta mag \sim5$).
It also requires the disk to have a high inclination (at least $60\deg$)
for forward-scattering by \micron-sized grains to be efficient at the gap outer edge.
We judge a disk-origin to be highly unlikely, since we detect a $CP$ as large as $30\deg$.

We calculated the probability $P$ for the $CP$ to be caused by contamination from a background 
source. $P$ is evaluated from the surface density 
$\rho(m)$ of sources brighter than $m= H +\Delta H = 9.1 + 2.5 = 11.6$ 
(worst case) around GG~Tau~A: $P(\Theta,m) = 1- \exp^{-\pi \rho(m) \Theta^2}$ \citep{brandner00}.
With 591 sources within a $\Theta \leqslant 1\deg$ radius circle around GG~Tau~A (after 2MASS, \citet{cutri03}) ,
$P$ is only $10^{-7}$ at a maximum separation of 0.032\arcsec.

Finally, re-examining archival NaCo data from December 2003 brings another 
argument in favor of a companion. 
We employed a procedure similar to that described in \citet{koehler00}: 
we extracted subframes around GG~Tau~Aa and Ab and applied speckle-interferometric 
techniques to compute the modulus of the visibility of Ab, using Aa as a PSF reference. 
The visibility shows clear signs of a close companion. 
A binary model was fitted, resulting in an Ab1--Ab2 separation of 
$28\pm3$\,mas, $PA = 65 \pm 5\deg$ [mod $180 \deg$], and a flux ratio of $4.6\pm1.0$ ($K_s$-band).
This {\it a posteriori} detection further supports the bound-companion
hypothesis. Assuming a circular orbit in the sky plane,
we derive a crude estimate of 
the orbital period $P = 16 \pm 1$\,yr for clockwise (CW) rotation from the 9\,yr separated positions.
Interestingly, this result agrees well with the orbital period derived according to 
the spectral type and stellar mass estimates (see Sect.\ref{Sec:triple} below), 
and CW rotation was also independently derived for the outer disk 
\citep[from the CO velocity field, ][]{guilloteau99} and for the Aa--Ab orbit.

\begin{table*}
\begin{tabular}{lcccccccccccc|}
\hline
\hline
    & \multicolumn{2}{c}{PIONIER} & \multicolumn{2}{c}{NACO-SAM} &  \multicolumn{2}{c}{NACO-SAM} &  \multicolumn{2}{c}{NACO-SAM} &  \multicolumn{2}{c}{ NACO-SAM} &  \\
    &  \multicolumn{2}{c}{$H$ (2012-10-30)} &\multicolumn{2}{c}{$H$ (2012-12-06/07)} &  \multicolumn{2}{c}{$K_s$ (2012-12-06)} & \multicolumn{2}{c}{$H+K_s$} &  \multicolumn{2}{c}{\Lprime\ (2012-12-07)} &   \\
    & best-fit & $\sigma$ & best-fit & $\sigma$ & best-fit & $\sigma$ & mean & $\sigma$ & best-fit & $\sigma$ \\
\hline
$\rho$ [mas] {\it Aa - Ab} & & & 256.1 & 1.5 & 255.8 & 3.5 & 256.1 & 1.3 & 256.3 & 3.3 \\
PA [deg]	{\it Aa - Ab} & & & 329.6 & 0.3  & 329.3 & 0.8	& 	329.5	& 0.4	 & 328.9	& 0.8   \\
F(Aa) / F(Ab)	& & &	2.36	& 0.04 & 1.99	& 0.04	&	& &  1.88 &	0.23	 \\
\hline
$\rho$ [mas] {\it Ab1 - Ab2} &31.7&	0.2 & 31.7	&4.1&	31.6&	5.8 &	31.6 &	3.3 & 	31.4&	fixed \\
PA [deg]	{\it Ab1 - Ab2} &219.6 & 0.3 &219.9	 &7.3 	&227.6	&10.3	&222.6	&6.0 &219.9	&fixed		\\
F(Ab1) / F(Ab2)		& $4.3^{(\ast)}$ & $2.3^{(\ast)}$ &1.6	& $^{+0.6}_{-0.2}$	&2.3	&0.2	&	& &2.8	&0.2	\\
\hline
\end{tabular}
\caption{Relative astrometric position and flux contrast for the main binary Aa-Ab (photo-center of Ab1--Ab2), and for the close binary Ab1--Ab2, as derived from our best-fit of the closure phases measured by VLT/NaCo-SAM and VLTI/PIONIER. The formal errors on NaCo-SAM $PA$ do not include uncertainties due the detector orientation. ($\ast$): after correction for FOV attenuation (see text), the related uncertainty dominates}
\label{table_astrometry}
\end{table*}

\onltab{
\begin{table*}
\begin{tabular}{lccccccccccccc|}
\hline
\hline
    & $A_V$  & \multicolumn{2}{c}{$I$} &  \multicolumn{2}{c}{$J$} &  \multicolumn{2}{c}{$H$}  &  \multicolumn{2}{c}{$K_s$} &  \multicolumn{2}{c}{\Lprime} &  \\
    &  & mag & $\sigma$ & mag & $\sigma$ & mag & $\sigma$ & mag & $\sigma$ & mag & $\sigma$  &\\
\hline
Aa (obs)      & 0.30 & 10.46 & 0.02 &  9.24 & 0.21 & 8.27 & 0.25 & 7.73 & 0.16 & 6.72 & 0.13&\\
Aa (dereddened) &          & 10.26 &          &  9.14 &         & 8.21  &         & 7.69  &          & 6.81 &      &\\
Aa (M0V)               &          & 10.26 &          &  9.26 &         & 8.63 &         &  8.46&           & 8.32 &&\\
\hline
$F_{\rm excess}$ / $F_{\rm tot}$ && 0      &           &   0.10 &  0.17&  0.32 & 0.16&  0.51 & 0.07& 0.75 & 0.03\\
\hline
Ab1+Ab2 (obs) & 0.45 & 12.19 & 0.04 & 10.12 & 0.21 & 9.07 & 0.32 & 8.53 & 0.24 & 7.69 & 0.04&\\
Ab1+Ab2 (dered.) &     & 11.89 &          &  9.97 &           & 8.98 &          & 8.47 &          & 7.66 &         &\\
Ab1 (M2V)           && 12.16   &          & 11.01 &         & 10.31 &        & 10.11&         & 9.95 &&\\
Ab2 (M3.5V)       &&13.53  &           & 12.13 &         & 11.48 &        & 11.23&         &11.02&&\\
\hline
$F_{\rm Ab1}^{\star}$ / $F_{\rm Ab2}^{\star}$ (M2/M3) && 2.1       &           &   1.9 &         &  1.9   &     &   1.8   &         & 1.7    &  \\
$F_{\rm Ab1}^{\star}$ / $F_{\rm Ab2}^{\star}$ (M2/M3.5) && 3.5       &           &   2.8 &         &  2.9   &     &   2.8   &         & 2.7    &  \\
\hline
$F_{\rm excess}$ / $F_{\rm tot}$ (M2/M3)&& 0       &           &   0.48 &  0.10&  0.61 & 0.12&  0.70 & 0.07& 0.83 &0.02\\
\hline
\end{tabular}
\caption{Photometry of the triple system: observed apparent magnitudes from \citet{hartigan03} for Aa and (Ab1+Ab2),
extinction-corrected magnitudes, expected magnitudes from color-indices and the spectral-type estimate,
expected flux ratios between stellar components for the closest spectral types, and estimated flux excess from circumstellar material for the adopted M2/M3V spectral types.}
\label{table:photometry}
\end{table*}
}

\onltab{\begin{table*}
\begin{tabular}{lcccccccccccc|}
\hline
\hline
{\it GG Tau Aa - Models}  & $\chi^2_{red}$ & $F_{halo}/F_{tot}$ & $F_{Ab1}/F_{tot}$ &   $F_{ring}/F_{tot}$  & $F_{excess}/F_{\rm tot}$ & $\Theta(ring) [mas]$ & R(ring) [au] \\
1- Point source+halo  & 1.4  & $0.10 \pm 0.01$ & $0.90 \pm 0.08$ &           --              &                           $\geq 0.11$      &      --         &       --          \\
2- Point source+halo+ring\,$^a$ & 1.2  & $0.06 \pm 0.01$ & {\it 0.68} (fixed)   &   $0.26\pm0.08$            &  {\it 0.32} (fixed)  &   $1.0 \pm 0.2$ &   $0.07 \pm0.01$ \\
\hline
\hline
{\it GG~Tau Ab1+Ab2 - Models} & $\chi^2_{red}$ & $F_{halo}/F_{tot}$ & $F_{Ab1}/F_{tot}$ & $F_{Ab2}/F_{tot}$  &  $F_{excess}/F_{\rm tot}$  & $(F_{Ab1}/F_{Ab2})_{\rm corr}$  \\
3- 2-point sources\,$^{b}$  & 1.5      &   --                          & $0.89  \pm 0.15$        &  $0.11\pm 0.02$            &       --    &      $4.3 \pm 2.0$  \\
4- 2-point sources+halo        &  2.2     &  $0.19 \pm 0.03$ & $0.73 \pm 0.09$        & $0.08 \pm 0.01$               &        $\geq 0.19$ & $4.6 \pm 1.8$  \\
\hline
\end{tabular}
\caption{Summary of geometrical model fits to the VLTI-PIONIER data on 2012-10-30 for GG~Tau Aa and Ab(1+2) respectively. 
($^a$): a circular ring and an elongated ring (with an aspect ratio, an inclination and orientation forced to the known circumbinary ring values) yield similar results. 
($^b$): fit to closure-phase values alone, a single-star model would yield $\chi^2_{\rm red} =7$.
}
\label{table_interferometry}
\end{table*}}

\subsection{Characteristics of the new triple stellar system} \label{Sec:triple}
If we assume that the close binary orbit is coplanar with the outer disk,
the de-projected separation of Ab1 and Ab2 was 5.1\,au in Dec. 2012.
Following the spectroscopic analysis of \citet{hartigan03},
the bright system GG~Tau~A is composed of Aa, an M0 star ($\sim 0.6$\,M$_{\odot}$),
and of Ab, which displays a M2V spectral type
($\sim 0.38$\,M$_{\odot}$ for a single star). 
\citet{hartigan03} noted that the inferred spectral type for Ab leads
to an insufficient stellar mass in the system in comparison with the dynamical mass estimate
of $1.28\pm 0.07$\,M$_{\odot}$ (for a distance of 140\,pc) derived from the CO gas kinematics in the ring
by \citet{guilloteau99}. We assume here that this spectral type M2 can be attributed to the brighter component Ab1.

If we assume that the relative flux excesses around Ab1 and Ab2 are roughly similar, 
the reported contrast approximates the stars contrast. 
Combining all flux ratio constraints in $H$, $K_s$ and \Lprime\ bands 
(see Table~\ref{table_astrometry}), and using the spectral type-color relations \citep{bessell91,leggett92}, 
we propose that Ab2 is an M3$\pm0.5$ dwarf star, with a mass $0.25-0.35$\,M$_{\odot}$ \citep{hartigan03} 
leading to a binary mass ratio $q_{Ab1-Ab2}=0.8 \pm 0.2$. 
If the orbit of Ab1--Ab2 is circular and coplanar with the outer ring, 
with a total stellar (dynamical) mass of $0.58-0.75$\,M$_{\odot}$ and 
a semi-major axis of $5.1\pm0.4$\,au, 
its orbital period is estimated to $P = 14 ^{+3}_{-2}$\,yr from Kepler laws. 
If the relative IR flux excess were much higher around Ab1 than Ab2, 
it would bias the spectral type estimate for Ab2 towards earlier types (i.e., $q\sim1$).

We note that a slightly better solution to the PIONIER $CP$ fit might be found
with a third component in the GG~Tau~Ab system, that is, if Ab2 were itself a close binary.
Such models are degenerated: several solutions exist with Ab2a--Ab2b separations 
of 10--15\,mas (1.5--2.3\,au, i.e., about a third of the distance
between Ab1 and the Ab2 barycenter), yielding $\chi^2_{\rm red}\sim1$ instead of 1.5 
($dof=27$), with a flux ratio $F_{\rm Ab2a}/F_{\rm Ab2b} \sim 1.5-3$. 
A significantly improved ($u,v)$ coverage would be necessary to decide for one of 
the two scenarios. 

Finally, the analysis of the PIONIER visibilities, in combination with an independent estimate 
of the excess flux around Aa and Ab from literature photometry, allows one to partly constrain 
the presence and characteristics of circumstellar material in the two systems. 
The related discussion can be found in Appendix~\ref{Sec:innerdisk}, 
where we show that Aa may be surrounded by a CS disk 
revealed by its marginally resolved bright inner rim, 
while the dust distribution around Ab1 and Ab2 might be more complicated.

\subsection{Consequences for the dynamics of the system}
In addition to solving the missing-stellar-mass problem, the discovery of a fifth component
in this system provides a logical explanation for the lack of submm/mm continuum emission around Ab
attested by PdBI \citep{pietu11} and ALMA observations \citep[submitted]{dutrey14}. 
The Ab1 and Ab2 components are indeed surrounded by Roche lobes of radius $\sim 2$\,au.
Tidal truncation effects naturally prevent the existence of stable disk(s) larger than this limit, 
and thus not massive enough to be strong mm emitters, in contrast to Aa. 
The 10$\mu$m silicate emission probably arises from warm grains located within 2\,au 
(i.e., well inside the Roche lobes). The resulting gravitational disturbance may also affect 
the supply of material from the outer regions through dust and gas streamers, 
as it breaks the symmetry of the Aa-Ab system. 
The close binary GG~Tau~Ab also strengthens the impact of tidal truncation 
onto the circumbinary ring. However, the location of its inner rim (180\,au), which is much more distant than
the apparent separation of the Aa and Ab(1+2) components, 
remains a puzzling characteristics of the system.

\begin{figure}
\centering
\includegraphics[width=5cm,angle=0]{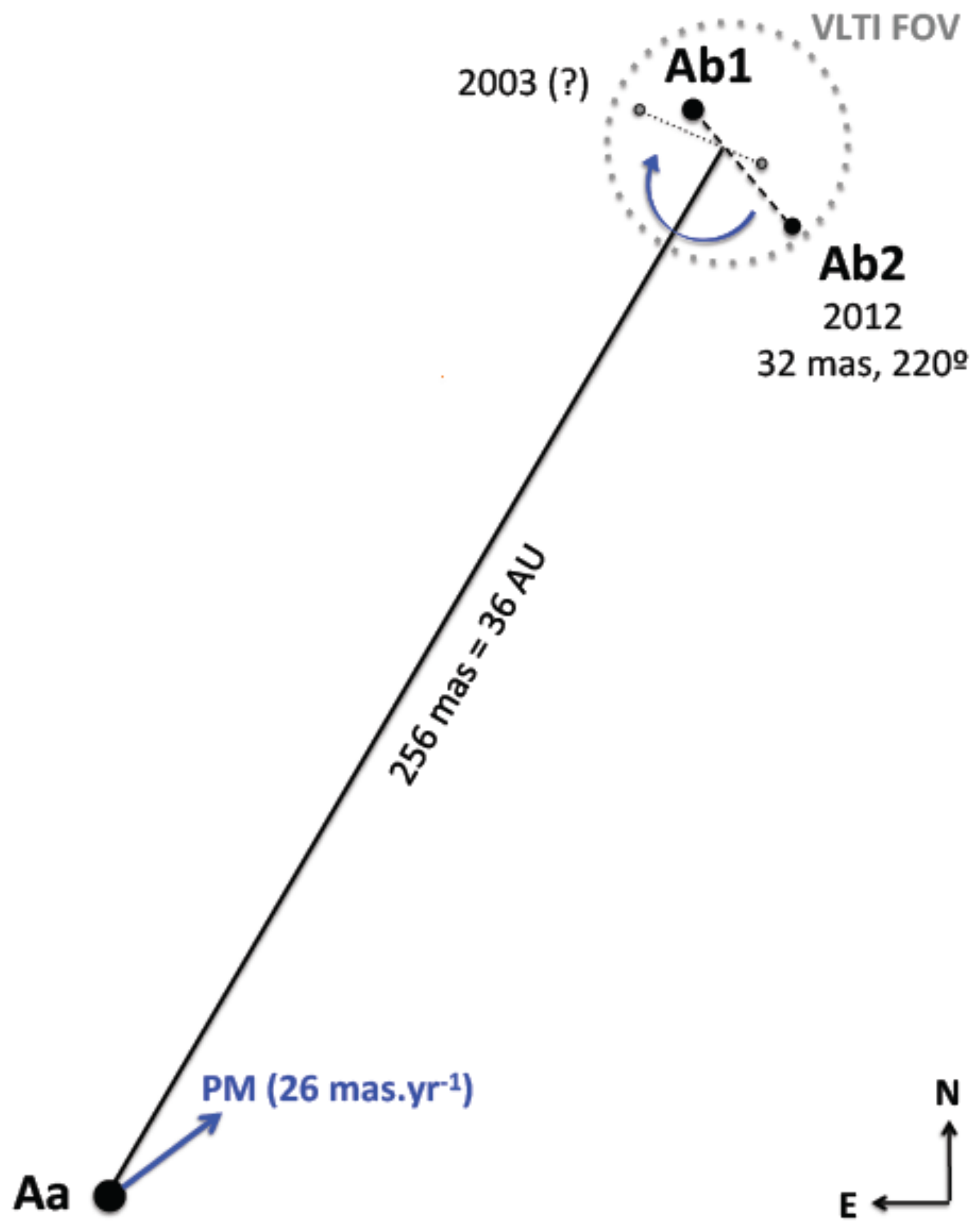}
\caption{Sketch of the new triple system GG~Tau~Aa (M0V), Ab1 (M2V), and Ab2 (M3V) in their 2012 orbital configuration. 
  }
  \label{fig_scheme}
\end{figure}
 
\section{Conclusion}
The emblematic, young binary system GG~Tau A (0.26\arcsec separation)
has been successfully observed with interferometric and
AO\,/\,sparse-aperture-masking techniques (SAM) at the Very Large Telescope.
We found that:

\begin{enumerate}
\item The secondary GG~Tau Ab is itself a close binary, with a projected
separation of $0.032\arcsec$ (or 4.5\,au)
and $PA=220^\circ$ (end 2012). It is consistent with a 
M3V (Ab2) and M2V (Ab1) low-mass binary.
This finding solves the discrepancy between the dynamical stellar mass 
derived from CO gas kinematics \citep{guilloteau99} and the most recent 
spectral-type estimate of Aa and Ab \citep{hartigan03}. 
Based on a tentative {\it a posteriori } identification in archival (2003) 
VLT/NACO images, its orbital period is estimated to $P_{\rm Ab1-Ab2} \sim16$\,yr,
a value consistent with the period derived from the binary mass and separation. 

\item All stars in this triple system present significant IR excesses, 
confirming the presence of circumstellar material. 
Around GG Tau Aa, the NIR emission is partly resolved at 1.65\micron\ and 
the derived geometrical ring radius is typical of proto-planetary disks around low-luminosity stars.
For GG Tau~Ab, due to tidal truncation, a (deprojected) separation of $\sim5.1$\,au sets
a strong constraint on the maximum radial extent of any circumstellar disk surrounding Ab1 and/or Ab2
(\Rout\,$\simless 2$\,au). The binary nature of this system also provides a simple explanation to the intriguing non detection
of mm continuum emission at the location of Ab.

\item With five coeval low-mass stars, this young multiple system becomes an ideal test case 
to constrain evolutionary models, provided that future astrometric studies 
will refine the stars physical parameters of the Ab system.
\end{enumerate}

\begin{acknowledgements}
We acknowledge the "Programme National de Physique Stellaire" and the "Programme National de Plan\'etologie" (CNRS/INSU, France) for  financial support. This research has made use of the Jean-Marie Mariotti Center SearchCal and LITpro services co-developped by FIZEAU and IPAG.
\end{acknowledgements}

\bibliographystyle{aa}
\bibliography{GGTau_Ab-VLT}

\Online
\begin{appendix}
\section{Inner disks constraints}\label{Sec:innerdisk}
Characterizing the inner circumstellar (CS) disk(s) is more challenging, 
because of the limited resolution of the interferometer. 
Not all system characteristics can be directly fitted. 
Using our spectral-type estimates for Aa and Ab1+2 and the 
resolved photometric values 
from the literature
 for Aa and Ab, 
we derived the relative contributions of the photospheres and of the CS material (IR excess), 
and use these inputs to model the VLTI data.

We first assumed that all the emission in the $I$ band is purely photospheric in origin 
and we adopted an optical extinction ratio $R_{V} = 5$ \citep[extinction laws from][]{mathis90}, 
and a common reddening value $A_V$ for Ab1 and Ab2 of 0.3 from \citet{hartigan03}. 
In the $H$ band, we derive a fractional excess emission  
$F_{\rm d} / F_{\rm tot} = 0.32 \pm 0.16$ for Aa,
and $F_{\rm d} / F_{\rm tot} = 0.61 \pm 0.12$ for Ab1+Ab2
(insensitive to the spectral type adopted for Ab2, see Table~\ref{table:photometry}).

We then fit the PIONIER data with analytical models that
consists of one (or two) star(s) plus a geometrically thin, circular ring to mimic 
the bright inner rim of the CS disk(s).  
For GG~Tau~Aa, 
a fully resolved component may contribute up to 7\,\% of the emission 
($V^2 < 1$ at short baselines, see Fig.~\ref{fig_v2+t3}). 
Such extended components (or "halos") are common around young stars, 
and can reach $\sim20$\,\% of the total emission \citep[e.g., ][]{akeson05}. 
It has been proposed that scattered light (at the disk surface or by a residual envelope) 
might explain this visibility drop \citep[e.g.,][]{monnier06,pinte08}.
Fixing the total excess flux ratio to 32\,\% (model-2 in Table~\ref{table_interferometry}), 
we infer a ring radius in the range $0.05-0.1$\,au, 
a value consistent with the expected grain sublimation distance
for a disk around a 0.38\,\Lsun\ star \citep{pinte08}.  
Our observation of Aa is thus consistent with an unresolved photosphere accounting for 68\,\%
of the $H$-band emission, surrounded by a canonical circumprimary disk 
whose inner rim bright edge remains marginally resolved by the VLTI. 
This disk is also known to be large and massive enough 
to produce detectable mm emission \citep{pietu11}. 

In the 32\,mas binary Ab, the presence of at least one CS disk is indirectly 
attested by the detection of the 10\micron\ silicate feature \citep{skemer11} 
and classical accretion tracers \citep{white99}. 
PIONIER visibilities do not show a clear drop at long baselines (Fig.~\ref{fig_v2+t3}), 
although the data are more noisy than for the brighter Aa. 
This indicates that any disk-like emission remains mostly unresolved. 
A fit of PIONIER data ($CP$+$V^2$) with two point sources and a halo-like component 
(see Table~\ref{table_interferometry}, model-4) 
yields $F_{\rm halo}$$\sim$20\,\%, $F_{\rm Ab1}$$\sim$70\,\% (star+ unresolved dust emission), $F_{\rm Ab2}$$\sim$10\,\%,  
and $F_{\rm Ab1} / F_{\rm Ab2} = 4.3\pm 2.0$. 
The discrepancy with the observed NaCo contrast ($1.6\pm0.4$) is hard to explain. 
We cannot exclude that it has an instrumental origin: 
the modest Strehl ratio $\sim 30$\,\% of the VLTI adaptive optics 
and tip-tilt correction residuals make the FOV correction delicate. 
However, if this discrepancy is real, it might be linked to the spatial extent and location of the halo emission in Ab. 
We propose that the halo might be located around Ab2,  
with a spatial extent in the range 10--30\,mas radius (i.e., 1.5--4.5\,au). 
It would thus be fully resolved on VLTI baselines, but would remain unresolved for NaCo, 
and would thus only contribute to the NaCo-SAM closure phase. 
This halo emission could partly originate from the complex geometry of streamers 
in the gravitationally unstable zone around the close binary. 
This is suggested by the detection of an extended warm $H_2$ emission around Ab, 
which most likely traces accretion shocks of inflowing material towards the CS disk(s) \citep{beck12}. 
If the halo and Ab2 emissions were co-located, the VLTI contrast (after correction for FOV-attenuation) 
would thus become $F_{\rm Ab1} / (F_{\rm Ab2}+F_{\rm halo})= 1.4 \pm 0.6$ (model-4), 
in better agreement with the NaCo-SAM value. 
Finally, from the photometry constraints (literature data, see Table~\ref{table:photometry}), we estimate that the $H$-band dust excess 
amounts to $60\pm10$\,\% of the total emission in Ab. Because Ab2 (star+disk) 
can only account for $\sim10$\,\% of the total flux, most of the remaining 40\,\% excess flux 
should arise from the Ab1 CS environment (disk?). 
Theoretical simulations of binary system formation \citep[e.g., ][]{bate97} 
suggest that the primary star usually accretes more material than the secondary, 
and in some cases the circumsecondary disk may not even be present. 
Although the limited spatial information and data quality do not allow us 
to fully constrain the CS environments in the Ab close binary, 
the current data set seems consistent with this scenario.

\end{appendix}

\end{document}